\documentclass[submission,copyright,creativecommons]{eptcs}
\providecommand{\event}{TFPiE 2024} % Name of the event you are submitting to

\usepackage{iftex}
\usepackage{graphicx}

\ifpdf
  \usepackage{underscore}         % Only needed if you use pdflatex.
  \usepackage[T1]{fontenc}        % Recommended with pdflatex
\else
  \usepackage{breakurl}           % Not needed if you use pdflatex only.
\fi

\title{Programming Language Case Studies Can Be Deep}
\author{Rose Bohrer
\institute{Worcester Polytechnic Institute\\ Worcester, MA, USA}
\email{rbohrer@wpi.edu}
}
\def\titlerunning{PL Case Studies Can Be Deep}
\def\authorrunning{Rose Bohrer}

\usepackage{amsmath,amsthm}
\usepackage{prettyref}
\newcommand{\rref}[1]{\prettyref{#1}}
\newtheorem{theorem}{Theorem}
\newtheorem{lemma}{Lemma}
\begin{document}
\maketitle

\begin{abstract}
In the pedagogy of programming languages, one well-known course structure is to tour multiple languages as a means of touring  paradigms.
This tour-of-paradigms approach has long received criticism as lacking depth, distracting students from foundational issues in language theory and implementation.
This paper argues for disentangling the idea of a tour-of-languages from the tour-of-paradigms.
We make this argument by presenting, in depth, a series of case studies included in the Human-Centered Programming Languages curriculum.
In this curriculum, case studies become deep, serving to tour the different intellectual foundations through which a scholar can approach programming languages, which one could call the tour-of-humans.
In particular, the design aspect of programming languages has much to learn from the social sciences and humanities, yet these intellectual foundations would yield far fewer deep contributions  if we did not permit them to employ case studies.
\end{abstract}

\section{Introduction}
The introduction is structured into four subsections, respectively addressing motivation (\rref{sec:intro-motivation}), context (\rref{sec:intro-context}), pedagogical philosophy (\rref{sec:intro-philosophy}), and contributions (\rref{sec:intro-contributions}).

\subsection{Motivation}
\label{sec:intro-motivation}
As programming languages (PL) educators, it is important to reflect on fundamental design decisions about course contents, the methodologies taught to students, and which intellectual traditions our courses will build on.
These decisions are important because, over time, they shape what the field of programming languages \emph{is} and  \emph{who} chooses to participate in the field, even perhaps shaping who does or does not feel they belong in PL classrooms.
In the author's classroom experience, many PL students are interested in the study of the \emph{design} side of PLs, yet find themselves with limited resources to study it.
The thesis of this paper is that when PL case studies are used to teach \emph{design} specifically, they provide irreplaceable contributions to PL pedagogy, compared to other uses of case studies in PL courses.
To support this thesis, the heart of the paper consists of a presentation of several key case studies from the author's \emph{Human-Centered Programming Languages} (HCPL) curriculum.

%This essay explores the structure of a programming languages (PL) course, specifically the potential that case studies hold for enriching a course. 
%Na\"{i}ve uses of case studies have been rightly criticized as intellectually \emph{shallow}; the mission of this paper is to reveal a non-na\"{i}ve use which is, on the contrary, intellectually \emph{deep}.
%This essay addresses the pedagogy of programming languages as a whole despite being addressed to an audience of functional programming educators.
%This is because the two audiences overlap heavily, and insights relevant to one are frequently relevant to the other.
%In both communities, we are deeply concerned with questions of foundations, of building up complex ideas from solid principles.
%In the big picture, this paper is an invitation to reconcieve our notion of ``foundations'' and incorporate a broad range of intellectual traditions in our pedagogy through the clever use of case studies, with the topic of design as a unifier.
%This invitation is equally relevant to both sides of this shared community.
%Before inviting a novel course structure, I begin with some history.
\subsection{Context}
\label{sec:intro-context}

This subsection provides context on the historical debate surrounding the role of PL case studies in PL pedagogy, to demonstrate why their use may be contentious to some readers.

Once upon a time, the gold standard of programming languages education was to take a tour of different programming paradigms by taking a tour of languages from each paradigm.
Imperative programming might be represented by C, object-oriented programming by Java, and logic programming by Prolog.
Functional programming languages could be represented by anything from a Lisp, to a member of the ML family (OCaml, Standard ML), to Haskell.
Yet there are vast differences between these languages: most Lisps are dynamically typed and strongly emphasize macros, the ML family is statically typed and impure with strict evaluation, and Haskell is statically typed and pure with non-strict evaluation.
The simple problem of identifying a PL to represent each paradigm reveals a major weakness in this course structure: paradigm labels do not neatly and usefully classify modern programming languages.
For example, the language Scala is undoubtedly functional and shares much with the ML tradition, featuring heavy use of type inference and type polymorphism, with strict evaluation.
At the same time Scala is undoubtedly an object-oriented language, with its Java interoperability a defining feature.
Conversely, Java is statically-typed with eager evaluation and now possesses first-class functions.
Yet, few functional programmers would declare Java to be a functional language, in part because the communities of Java programmers vs.\ Scala programmers have radically different cultural traditions.
It has become clear with time that the notion of classifying languages by paradigms falls apart under any close analysis, one of many reasons this approach has waned in popularity over the years.

One of the clearest criticisms of the tour-of-languages approach is Shriram Krishnamurthi's essay \emph{Teaching Programming Languages in a Post-Linnaean Age}~\cite{DBLP:journals/sigplan/Krishnamurthi08}, which advocates instead for a \emph{tour-of-features} approach, embodied by his textbook \emph{Programming Languages: Application and Interpretation} (PLAI)~\cite{DBLP:books/daglib/0017450}.
In this approach, a lesson might address a feature such as polymorphism, Hindley-Milner type inference, or lazy evaluation, but never a specific real-world language that exhibits these features.
Courses using this approach often assign the implementation of each feature as classwork.
That implementation often takes place in a Racket-based environment, writing interpreters for a pedagogical language with a Lisp-style syntax.
The textbook contains ample example programs using each feature, and writing such programs can be included as part of coursework.

To his credit, Krishnamurthi acknowledges the tradeoffs inherent in each approach, which reflect a tension between inductive and deductive learning styles.
The use of concrete existing programming languages can help students learn inductively from pre-existing example languages, while a tour-of-features places greater emphasis on deductive learning from the start, then adds example programs in the implemented language in an effort to meet both learning styles.

\subsection{Pedagogical Philosophy}
\label{sec:intro-philosophy}
The subsection characterizes the pedagogical philosophy of the HCPL project, of which this paper is one part.
We contrast different course philosophies as different kinds of ``tours'': the \emph{tour of features} (PLAI), the \emph{tour of paradigms} (pre-PLAI), and the \emph{tour of humans} (HCPL).
Not all texts are organized as tours, and many~\cite{harper2016practical,pierce2002types,pierce2004advanced,reynolds1998theories}  emphasize theory to the exclusion of other topics. 
Though such texts remain cornerstones of the field, it is essential not to define competing intellectual foundations as ``outside PL''; given PL's status as a prestige subfield~\cite{laberge2022subfield}, such rhetoric serves to illegitimize such competing approaches.

The HCPL curriculum simultaneously owes a great debt to PLAI's \emph{tour-of-features} approach yet aims to overcome a substantial limitation of pre-existing approaches in general.
In particular, the designers of most contemporary curricula~\cite{friedman2001essentials,harper2016practical,DBLP:books/daglib/0017450,mitchell1996foundations,nystrom2021crafting,pierce2002types,pierce2004advanced,reynolds1998theories}, PLAI included, recognize both the importance of PL \emph{design} and the importance of ``teaching to the 90\%,'' working to define a broad potential audience for a PL course.
I wish to argue that these noble goals, which HCPL shares, fundamentally contradict the move away from case studies represented by the \emph{tour-of-features}.
Case studies are at the heart of any intellectually serious analysis of \emph{design}, whether that analysis finds its foundations in the social sciences or in the humanities.
In our efforts to ensure an intellectually serious treatment of the mathematical foundations of PLs, we must not forget that design deserves this same seriousness.

To treat both the mathematics and design of languages with proper respect in a single course is no small feat, yet case studies are the key ingredient that makes this possible, serving as the basis of the \emph{tour-of-humans} philosophy.
The HCPL curriculum is organized as a tour of different intellectual traditions for  PLs, i.e., a \emph{tour of the different humans who study PLs}.
This philosophy is inseparable from HCPL's use of case studies: the lesson surrounding each case study explores how specific intellectual traditions would engage with the design of each language.
The fundamental challenge of the HCPL project is that it must cover much breadth and sacrifice depth; the tour-of-humans helps maintain a coherent course structure while moving from case study to study, all while putting scholars of proofs and of humans on equal footing.
Without case studies, covering this breadth in a single course would be daunting.

Moreover, the human-centered side of PLs cannot be studied without attention to actually-existing languages as they are actually used, and thus cannot be studied properly in the absence of case studies.
Because HCPL also incorporates foundations and implementation, however, these case studies do not serve to eliminate rigor, instead reinforcing the foundations of programming languages while also integrating with foundational techniques in the social sciences and humanities.
For detailed information on the connections drawn and the intellectual depths of each case study, proceed to \rref{sec:case-studies} and consult \rref{tab:depth-table}.

\subsection{Contributions}
\label{sec:intro-contributions}
The core contribution of this paper is a detailed description of five case studies from the Human-Centered Programming Languages (HCPL) curriculum, detailed in \rref{sec:case-studies}.
Though the same underlying case studies appear in the HCPL textbook~\cite{hcpl}, the presentation is thoroughly different.
The textbook addresses each case study in a pedagogical style designed for student consumption, whereas the present paper deconstructs each case study for an audience of PL educators, pulling back the curtain to reflect on how each one fits into the curriculum.

Secondarily, this paper contributes argumentation and reflection, assembling the case studies to make a broader argument about the potential role of case studies in PL pedagogy and reflecting on the fluid, evolving definition of the field of PL as a whole.

\paragraph*{Positionality Statement}
I discuss my positionality to provide context to the analyses present in this work.
The HCPL project addresses issues of gender and disability as they interact with PLs.
The treatment of these topics is informed by experience as a binary transgender woman with multiple disabilities, both physical and developmental: Ehlers-Danlos Syndrome and neurodivergence.
Identity is not a claim to authority, but lived experience is a valuable resource in interpreting academic scholarship and synthesizing it into a curriculum.
Academic background also provides useful context: the author's PhD is in theorem-proving with substantial elements of PL theory, yet she has increasingly added both human-computer interaction and CS education research to her research agenda. Thus, the HCPL curriculum is developed from the perspective of a Computer Scientist working outward to connect with other disciplines.

Moreover, the goal of inclusive pedagogy will never be achieved by centering any single person with a marginalized identity, but only through a diversity of perspectives that can draw on our community's wide range of intersecting identities.
Readers should be aware of other marginalized identities whose life experiences are essential to a complete HCPL curriculum, yet the author cannot speak to the experiences of these identities.
Notable examples are native language, race, and socio-economic status.
The author is a native-English-speaking, white American employed as a tenure-track professor, with the socio-economic privilege that provides.
Of these factors, it is established in the literature that native language is an important factor in inclusive PL design~\cite{DBLP:conf/splash/SwidanH23}.
Though I am unware of specific publications on the latter topics, racism and socio-economic exclusion are common topics of informal conversation with my colleagues in the context of the PL community and higher education more broadly.
I call for the increased publication of studies which can draw on these life perspectives and other intersectional identities that the author lacks.

I outline the remaining sections.
Related work is in \rref{sec:related-work}.
Background information on the design of the HCPL curriculum is given in \rref{sec:structure}.
The heart of the paper is \rref{sec:case-studies}, which presents the case studies.
The conclusion is \rref{sec:conclusion}.

\section{Related Work}
\label{sec:related-work}

\paragraph*{Human-Centered Programming Languages}
This work is part of the broader \emph{Human-Centered Programming Languages} project~\cite{hcpl,DBLP:conf/splash/Bohrer23}.
In addition to the book itself~\cite{hcpl}, an initial companion paper describes the principles behind the book's construction, the origins of the archetype-based approach, book contents, and insights from data analysis of course reports~\cite{DBLP:conf/splash/Bohrer23}.
Though that paper lists the case studies from the book, they are not documented in depth.
The integration of Rust in the HCPL course is informed by the same author's prior Rust curriculum proposals~\cite{IntroRust}.

In contrast, this paper focuses exclusively on the HCPL case studies. 
It is our goal to supplement the textbook itself by giving educators a clear picture of how the case studies can be used in the classroom, what kinds of activities might be used in such lessons, and how they fit together with the rest of the course.
In doing so, this paper also serves to argue that these case studies can be used to reinforce the fundamentals of design especially, and even the fundamentals of PL theory.
By design, this paper recounts the specific case studies of HCPL at great length, with the aim of arguing this broader claim.

\paragraph*{User-Centered PLs and Programmer Experience}
The HCPL project is far from the first study of the relationship between humans and programming languages.
A substantial body of work exists under two labels: \emph{User-Centered PLs}~\cite{DBLP:journals/computer/MyersKLY16} and \emph{Programmer eXperience} (PX)~\cite{DBLP:journals/access/MoralesRBQ19}, with the latter label emphasizing the inclusion of tooling beyond the language definition itself.
Notable works within this area include the Whyline~\cite{DBLP:conf/chi/KoM04}, a debugger with explanations, and the study of natural programming~\cite{DBLP:journals/cacm/MyersPK04}, i.e., based on  programming user studies where the subjects have no prior conceptions of programming to bias their experiences.
In recent years, this line of work has been operationalized in the PLIERS design framework~\cite{DBLP:journals/tochi/CoblenzKKWBSAM21}.

The distinguishing feature between HCPL and (the majority of) User-Centered PL and PX approaches is that the latter address the human element of programming primarily through the notion that a programmer is a user of a software system for the task of programming.
As a result, these approaches are rooted in HCI, more specifically in user experience (UX).
In contrast, the focus of HCPL is on the breadth of different humans who study PLs, the archetypes, which also distinguishes us from the wide array of existing high-quality PL textbooks
~\cite{friedman2001essentials,harper2016practical,hermans2021programmer,DBLP:books/daglib/0017450,mitchell1996foundations,nystrom2021crafting,pierce2002types,pierce2004advanced,reynolds1998theories}.
This intentionally includes not only perspectives on usability, but critical perspectives from the humanities, while also putting these views into conversations with approaches to PLs rooted solidly in theoretical computer science.
That said, the perspectives represented in the User-Centered PL and PX communities are not homogenous, and some authors' perspectives align more closely with that of HCPL.
For example, Ko~\cite{DBLP:conf/oopsla/Ko16a} invites viewing PLs as socio-technical systems; as is HCPL, this is fundamentally an invitation to a highly interdisciplinary perspective.

\paragraph*{Pedagogy}
The design of HCPL is informed by a number of concepts in pedagogy scholarship.
A main strategy for helping students navigate the technical breadth of the course is by providing flexibility and transparency in the evaluation process.
An early iteration of the course took a fully project-based~\cite{kokotsaki2016project} approach, before pivoting to a blend of autograding and completion\footnote{Completion grading is a grading scheme where full points are awarded on all problems that are completed, irrespective of any quality assessment} grading with peer-review elements in subsequent instances of the course.

These pedagogical design decisions are informed by the philosophy of ungrading~\cite{stommel2018ungrade,kohn2020ungrading,ferns2021ungrading}, which views rating and ranking of students as fundamentally detrimental to the process of learning, aiming instead to maximize intrinsic motivation, often through choice and ownership.
In ungrading contexts, a common challenge is student self-regulation~\cite{zimmerman2002becoming}, a trait which is associated with future student success but can rarely be cultivated by a single course.
In response to this challenge, the present iteration of the HCPL course uses autograding as a means of providing external regulation of students, under the hypothesis that once students have fulfilled one set of classwork obligations thanks to external regulation, they are more likely to fulfill their completion-graded assignments as well.
See prior work~\cite{DBLP:conf/splash/Bohrer23} for further discussion of the relationship between HCPL and ungrading, including student reactions to the practice, both positive and negative.

HCPL is not the only open-access PL textbook nor the only to provide open-access companion materials, but the choice of open access was an intentional one.
Open education~\cite{giaconia1982identifying,hylen2006open,seely2008open,wiley2014open} is the long-running movement to make educational as widely available as possible through the reduction of barriers.
The use of open access and even, in part, the use of autograding are intended to enable an open education experience, e.g., for self-study outside a university setting.

The HCPL course is potentially challenging to execute for instructors because of its technical breadth.
One strategy for coping with its breadth is to keep an explicit focus methodology, and bring students back to the different methodologies used.
In written tasks from completion-graded assignments, the literatures on reflective journals~\cite{george2002learning} and autoethnography~\cite{ellis2011autoethnography} can both guide classroom practice.
In the social science modules of HCPL, key techniques include the use of thematic analysis~\cite{braun2012thematic,clarke2015thematic} and personas~\cite{DBLP:series/hci/Nielsen13}; ample resources are available for both.

Other teaching methods in HCPL have their origins in the humanities and have seeped into CS over time.
In particular, the use of dialogues between archetypes has thousands of years of methodological precedent in the form of Socratic dialectic~\cite{delic2016socratic,benson2006plato}, though our dialogues by comparison are intended to reveal a multitude of co-existing perspectives instead of guiding a student to a specific truth known by a teacher.
In the modern era, \emph{Proofs and Refutations}~\cite{lakatos1963proofs} is a well-known use of dialectic to teach computer scientists about mathematical proof, which is at the heart of PL theory.
Educators firmly within the humanities can dive into more specific methodologies not addressed in depth by the HCPL text, such as codework~\cite{codework}, poetry in the medium of code.

The HCPL approach is also indebted to the literature of PL pedagogy specifically, not limited solely to Krishnamurthy's essay on the role of paradigms in PL education~\cite{DBLP:journals/sigplan/Krishnamurthi08}.
For example, recent scholarship on social issues education for CS students has advocated integration across the curriculum instead of standalone courses~\cite{ResponsibleComputing}; HCPL's coverage of gender and disability issues implements this approach.
The use and discussion of toy teaching languages in HCPL is informed by the lessons learned by other PL education researchers; classroom discussions of the transition from a teaching language to a production quality language are partially inspired in particular by \emph{A Data-Centric Introduction to Computing}'s~\cite{dcic} explicit coverage of this transition process.

Like any new course design, the development of HCPL was motivated by the dearth of existing courses that filled its interdisciplinary niche. At the same time, the few prior courses that filled this niche are thus all the more important.
In particular, the work of Michael Coblenz, Jonathan Aldrich, and their collaborators~\cite{DBLP:journals/corr/abs-2011-07565} directly influenced the first iteration of the HCPL course, an influence which thus carries through indirectly in the book.

\paragraph*{Case Studies}
Most of the case studies discussed herein are already documented in the research literature.
I cite this prior documentation, which is the basis of my own work, respectively for Penrose~\cite{DBLP:journals/tog/YeNKMWASC20}, Torino~\cite{DBLP:journals/hhci/MorrisonVTATSCS20},
FLOW-MATIC~\cite{hopper1959automatic,flowmatic}, Processing~\cite{reas2006processing}, and Twine~\cite{Twine}.

The C-Plus-Equality case study is the exception; I am not aware of any prior peer-reviewed literature engaging with this case study, and minimal scholarly writing about it any form~\cite{notLaughingCPE}.
By helping bring this case study into the literature, I hope to raise awareness among PL researchers of gendered harassment that has used the vocabulary of PL.
To highlight the importance of documenting this case study: the full case study materials were publicly visible on GitHub~\cite{cpeRepo} when I began preparation of this manuscript, but the repository was disabled by GitHub\footnote{The repository was disabled for terms-of-service violations due to its misogynistic content. My analysis aims to shed light on that misogyny.} by the time of submission.
Beyond this specific case study, the discussion of gender in HCPL has been informed by various scholars~\cite{mezangelle,DBLP:journals/iwc/BurnettSMMBKPJ16,disturbance,hastac}.

\section{Course and Book Structure}
\label{sec:structure}
This paper is dedicated to the case studies of the HCPL curriculum, arguing for their pedagogical validity and documenting them in the process.
For detailed documentation on the HCPL course, see prior work~\cite{DBLP:conf/splash/Bohrer23}.
Here, we provide relevant background information about the course's contents in \rref{sec:contents}, its use of \emph{archetypes} in \rref{sec:archetypes}, and the coursework in \rref{sec:coursework}.

\subsection{Contents}
\label{sec:contents}

I list the chapter titles, length estimates\footnote{The length estimates are provided automatically by the publishing platform, Bookish.press. Reading time varies by reader, but these estimates give a sense of relative length.}, and languages discussed in \rref{tab:table-of-contents}.
Many chapters contain topics which could be viewed as case studies; I list which case studies were included in this paper and briefly discuss how they were selected.
\begin{table}
\caption{Table of contents for current iteration of book}
\begin{tabular}{ccccc}
No. & Title & Length & PLs & In Paper?\\
1 & Introduction & 10min &  & \\
2 & What is a Language & 25min & & \\
3 & Programming in Rust & 1hr & Rust & N\\
4 & Regular Expressions & 35min & RE & N\\
5 & Context-Free Grammars & 50min & CFG & N\\
6 & Parsing Expression Grammars & 30min & PEG & N\\
7 & ASTs and Interpreters & 35min & & \\
8 & Operational Semantics & 5min & &\\
9 & Types & 50min & &\\
10 & Users and Designers & 40min & & \\
11 & Quantitative Methods \& Surveys & 45min & Randomo & N \\
12 & Qualitative Studies & 50min & & \\
13 & Gender & 35min & C+= & Y\\
14 & Disability & 30min & Torino & Y\\
15 & Media Programming & 15min & Processing & Y\\
16 & Play & 15min & Twine & Y\\
17 & Natural Language & 25min & FLOW-MATIC,Inform & Y,N\\
18 & Diagramming & 15min & Penrose & N\\
19 & Process Calculus & 20min & Proc.\ Calc. & N\\
20 & Cost Semantics & 20min & Par.\ ML & N
\end{tabular}
\label{tab:table-of-contents}
\end{table}

Given the interdisciplinary nature of the course and textbook, I am often asked how the book's contents align with more traditional intradisciplinary courses.
My courses are housed in a CS department at an institution which does not segregate HCI from CS, thus it contains substantial elements of both CS (Ch.\ 1--9) and HCI (e.g., in Ch.\ 10--14 and 18).
For instructors concerned about integrating an interdisciplinary course into a disciplinary setting, one potential strategy is employing a subset of the HCPL text in concert with other materials.

As \rref{tab:table-of-contents} shows, not all of the HCPL case studies are covered in this paper.
I used the following considerations in choosing which case studies are presented in the current paper. 
I excluded case studies where I believed there is substantial debate (e.g. between the archetypes) about whether it \emph{is a case study}, such as PEGs and process calculus.
I excluded case studies where writing about them would substantially duplicate another case study, such as Inform and Penrose.
Lastly, I prioritized case studies with little existing documentation in the academic literature by including C+= which is ill-documented and excluding Randomo, which is well-documented.

\subsection{Archetypes}
\label{sec:archetypes}
In arguing our thesis that PL case studies can be deep, an awareness of course structure is essential.
The HCPL course structure is called the \emph{tour-of-humans} structure, where students tour between different \emph{archetypes} representing different scholarly perspectives.
The five archetypes are named the Practitioner, Implementer, Theorist,  Social Scientist, and  Humanist. 
I give a subjective listing of the chapters in which each archetype takes a substantial role: Practitioner (1--3,15--18\footnote{Lessons 15--18 involve classroom coding exercises in case study languages}), Implementer (4--7), Theorist (8,9,16,19,20), Social Scientist (10--14,18), Humanist (13--15,17).
In the text, the archetypes are used to explain motivations and approaches to solving problems.
They are put into conversation with one another using dialogues within certain chapters.

To contextualize the overall HCPL course design, I summarize each archetype.
I do so by characterizing who they are, what questions they ask, how students interact with them, and how the archetypes relate to the case studies.

\textbf{The Practitioner:}
The Practitioner is someone who interacts with code as a programmer, but does not implement, design, nor theorize about programming languages. 
The Practitioner's fundamental question is "How do I write this program?"
Students engage with this archetype by writing code and reflecting on their experience, including classroom activities with case study languages.

\textbf{The Implementer:}
An Implementer is anyone who implements a programming language, typically as a compiler or interpreter. 
The Implementer’s fundamental question is “How do I implement this programming language?”
Students engage with this archetype by implementing PLs in coursework.

\textbf{The Theorist:}
The Theorist is anyone who does the work of defining PLs as  formal languages or  analyzing them mathematically. To them, a “good PL” is a language that we can analyze in powerful ways. A Type Theorist is a Theorist who believes  a “good PL” has a rich static type system that lets us prove powerful theorems about the correctness of programs. 
%Examples of books the Theorist might read are “Types and Programming Languages” by Benjamin Pierce and Concrete Semantics by Gerwin Klein and Tobias Nipkow.
The Theorist's fundamental question is "What can I prove about this language?"
Based on my students' needs, I limit engagement with the Theorist to reading and contextualizing proofs and mathematical notations.
Case studies (e.g., Twine) engage with the Theorist by highlighting practical applications of theory.

\textbf{The Social Scientist:}
The Social Scientist is someone who undertakes rigorous academic study of humans. 
 %Examples of books the Social Scientist might read include “Working in Public: The Making and Maintenance of Open Source Software” by Nadia Eghbalee. 
 The Social Scientist’s fundamental question about programming languages is “How do programming languages affect communities of people?”
Students engage with the Social Scientist by designing self-directed user studies and performing them on classmates.
The case studies engage with the Social Scientist both by teaching social science methodologies (Randomo) and by showcasing their potential contributions to novel designs (Torino, Penrose).

\textbf{The Humanist:}
The Humanist also studies humans, thus I am often asked how and why HCPL distinguishes the Humanist from the Social Scientist.
Both archetypes are employed in HCPL because PL design can benefit from motivations and methodologies firmly within the humanities, particularly the use of critiques driven by careful textual analysis.
As with all the archetypes, real-world Humanists are more varied than a single character can portray.
For our HCPL archetype, however, I restrict the Humanist's fundamental question to be ``How can social analysis be applied to PLs?''
Students engage with the Humanist through a self-reflection exercise drawing on their own experience and by following lectures which employ humanistic techniques.
Case studies engage with the Humanist through the use of historical methods (FLOW-MATIC) and hermeneutic inquiry (C+=), and by drawing on the model of the disability spectrum from disability studies (Torino).

The Social Scientist and Humanist receive particular focus in this paper; for deeper discussion of the roles of the other archetypes, see prior work~\cite{DBLP:conf/splash/Bohrer23}.

%It asks who communities include or exclude? Why? What could be done about that? The difference is that it uses methods from the humanities. Important books about computing could be read closely and their language analyzed. People look at rhetoric about languages and rhetorical structure present in code itself. People do theory-building, taking core ideas from social theorists and applying them to the specifics of PL communities.
 
%Examples of books the Humanist might read are “Persuasive Games: The Expressive Power of Videogames” by Ian Bogost and “Rhetorical Code Studies” by Kevin Brock. 

\subsection{Coursework}
\label{sec:coursework}
Successful pedagogy with the HCPL approach hinges on flexible design of coursework.
As of this writing, the course uses a ``\emph{two-spoke}'' model, where the first spoke consists of auto-graded programming assignments and the second consists of completion-graded written assignments with student peer-review. As of the Fall 2023 iteration of the course, the programming assignments prioritize the Implementer and the written assignments prioritize primarily the Social Scientist, followed by the Humanist.
The Theorist is assessed indirectly in programming assignments which rely on understanding of parsing, type systems, and evaluations, but receives the most focus in exam questions that directly address theory.

The current course iteration has five assignments, each divided into a programming and written assignment with the same deadline.
The programming assignments are: a Rust language warmup, a parser, an interpreter, a type-checker, and a small library of quantitative data analyses.
The respective written assignments are: a questionnaire on students' personal learning goals, an initial brainstorm for student-directed PL user studies, a reflection on personal experiences of Rust, a full user study proposal, and a writeup of user study results.

What is the relationship between the coursework and the case studies?
On the one hand, these assignments do not require knowing case study details---the Fall 2023 course assesses factual case study knowledge through a final exam instead.
However, the course's core social science skills are refined throughout the lectures by their repeated appearances throughout the case studies, and the written assignments fundamentally require these skills. Though case study details are not tested on homeworks, the case studies still have the critical role of preparing students for their homework.
It is natural for students to wonder how students engaged with written homeworks given that they are graded by completion and that the design material may be very new to many students.
In the Fall 2023 undergraduate course, students showed substantial creativity, such as the development of text adventure games to employ in their user studies, bringing musical instruments to class to explore musical PL interfaces, and physical activities to test the role of posture in programming.
These classroom experiences are consistent with the common assertion by \emph{ungrading}~\cite{stommel2018ungrade} proponents that creativity increases when grading is removed but intrinsic motivation remains.

\section{The Case Studies}
\label{sec:case-studies}
We now arrive at the heart of the paper.
In this section, I present the HCPL case studies.
For each case study in turn, I argue the thesis that these case studies reinforce foundational concepts in support of key learning goals, each with specific design implications
Colloquially, I argue that PL case studies can be deep.

In \rref{tab:depth-table}, I list out each case study, the topic(s) it connects with, and its \emph{depth}, i.e., which concept or methodology is fundamental to the case study analysis, each with a design implication.

\begin{table}[!htbp]
    \centering
    \begin{tabular}{c|c|c|c|c|c}
       Case        & FLOW-MATIC &  Processing & Twine & Torino & C+= \\
       Connection  & History    &  Media Arts   & IF        & Disability Stud.\ & Gender Stud.\ \\
       Depth       & Sources    &  Them.\ Anal. & FSMs      & Disab.\ Spectrum  & Hermeneutics \\
       Implication & Audience   &  Continuity   & Winnability & Hybrid Syntax &  Commun.\ Discourse
    \end{tabular}
    \caption{Connections, Depths, and Implications of Case Studies}
    \label{tab:depth-table}
\end{table}

The FLOW-MATIC case study connects with the field of history, its depth lies in collection and interrogation of primary sources. Its design implication is that FLOW-MATIC's designers did identify a specific audience, enabling us to apply contemporary audience-driven analyses, despite the fact that modern design methodologies had not yet been formalized at that time.
The Processing case study connects on the surface with the field of media arts, but its methodological depth is the social science method of thematic analysis; throughout this lesson I show how a complex mix of written text, visual artifacts, and lived experiences can be used to form our opinions as  designers, going beyond the more basic or rigid forms of thematic analysis.
A key design implication is that the notion of continuity between languages is two-sided, and designers ought to be aware of which languages might be learned \emph{after} a programmer has learned their language.
The Twine case study connects on its surface with the topic of interactive fiction (IF), yet its methodological foundations are solidly within CS theory: finite state machines (FSMs).
The direct design implication is that Twine's structure enables static analysis of Twine games' winnability, yet the broader learning implication is that CS theory might be applied in surprising places.
The Torino case study connects with the field of disability studies and its depth lies in its conception of disability as a spectrum.
The design implication is that hybrid syntaxes (e.g., tactile-and-visual or aural-and-textual) support collaboration between programmers across multiple positions on the disability spectrum.
The C+= case study connects with the field  of gender studies and its depth lies in its hermeneutic inquiry~\cite{schleiermacher1998,smith1991hermeneutic} approach, which embraces the notion of textual interpretation as an active process of  creation.
The design implication of the case study is that a designer cannot fully understand the user experience of their language without careful attention to ongoing discourse between members of the language community, including elements of discourse that must be carefully read out of the implications of a document.

We now proceed to the  case studies: FLOW-MATIC (\rref{sec:cs-cobol}), 
%Inform (\rref{sec:cs-inform}), 
Processing (\rref{sec:cs-processing}), Twine (\rref{sec:cs-twine}), 
%Process Calculus (\rref{sec:cs-pc}), Penrose (\rref{sec:cs-penrose}), 
Torino (\rref{sec:cs-torino}), and C+= (\rref{sec:cs-cpe}).

\subsection{FLOW-MATIC}
\label{sec:cs-cobol}
This case study is an invitation to students to the study of the history of PLs.
To the Humanist, the work of history is often the work of reconstructing an image that will never be complete, building narratives and arguments about the past based on the small window provided by the historical record.
In the history of PLs, we often forget to teach this element, perhaps because our own history is so brief, or perhaps because we are not professional historians.

The case study on FLOW-MATIC~\cite{flowmatic} (predecessor of COBOL) seeks to expose PL students to the work of history, of reconstruction.
This case study is introduced shortly after students are exposed to fundamental contemporary design tools, such as the use of user personas, design of user studies, and interpretation of qualitative data from user studies.
In approaching the history of PLs, we are confronted with the fact that most well-known PLs, both historically and today, did not follow any such formal design process.
This does not mean that their creators lacked all design knowledge, but rather that the process was far less formalized and recorded far less thoroughly than it might be in a modern approach.
The goal of this case study, then, is making sense of surviving documents from early PL designs and extracting an understanding of their design goals and approaches that are comprehensible to a modern audience.
FLOW-MATIC was chosen in particular because it is the earliest known PL which sought to mimic natural language and because connections with natural language remain a topic of curiosity for students to this day.
Thus, our research questions as historians of FLOW-MATIC are:
``What goals and context motivated the earliest natural-language PL design efforts?
What is the legacy of those efforts, and how can this legacy inform the role of natural language in contemporary design?''

The first primary sources for this historical analysis are the recorded statements of Grace Hopper, who was the designer of FLOW-MATIC and a core figure in the COBOL design process. The following quote~\cite{hopper1959automatic} from Hopper is used to contextualize the design goals and motivation of FLOW-MATIC:

\begin{quote}
I used to be a mathematics professor. At that time I found there were a certain number of students who could not learn mathematics. I then was charged with ``the job of making it easy for businessmen to use our computers''. I found it was not a question of whether they could learn mathematics or not, but whether they would. [...] They said, ‘Throw those symbols out; I do not know what they mean, I have not time to learn symbols’
\end{quote}

This quote serves as an opportunity to show a direct connection between foundations and case studies. A previous foundational lecture uses the ISO 9241-11~\cite{ISOUsability} definition of usability to teach students what it means to define a usability problem and assess a proposed solution. Students are taught to consider three questions of problem definition and three of assessment, respectively: defining users, goals, and context, and assessing effectiveness, efficiency, and satisfaction.

In teasing apart Hopper's statement, I show that it addresses elements of usability that are echoed by a standard released decades later.
She provides a clear definition of the user population: businessmen (and, in the full statement, military staff), which carries with it implications about goals and especially context. Her quote reflects the idea that satisfaction was the root concern and that effectiveness had been hindered by dissatisfaction; perhaps efficiency was not the top design priority. The unwillingness to use symbols is also related to discussions of programmer self-efficacy, which arises also in the lecture on gender.

Our second primary source gives students a chance to engage hands-on with specific code.
Official, contemporaneous documentation of FLOW-MATIC~\cite{flowmatic} has been well-preserved by historians of computing, including source code of example programs.
The following program text is presented in the classroom and students are given time to read it and guess at its meaning before further instruction:
\begin{verbatim}
INPUT INVENTORY FILE-A PRICE FILE-B ; OUTPUT PRICED-INV FILE-C UNPRICED-INV
  FILE-D ; HSP D .
1  COMPARE PRODUCT-NO (A) WITH PRODUCT-NO (B) ; 
   IF GREATER GO TO OPERATION 10 ;
   IF EQUAL GO TO OPERATION 5 ; OTHERWISE GO TO OPERATION 2 .
2  TRANSFER A TO D .
3  WRITE-ITEM D .
4  JUMP TO OPERATION 8 .
5  TRANSFER A TO C .
6  MOVE UNIT-PRICE (B) TO UNIT-PRICE (C) .
7  WRITE-ITEM C .
8  READ-ITEM A ; IF END OF DATA GO TO OPERATION 14 .
9  JUMP TO OPERATION 1 .
10 READ-ITEM B ; IF END OF DATA GO TO OPERATION 12 .
11 JUMP TO OPERATION 1 .
12 SET OPERATION 9 TO GO TO OPERATION 2 .
13 JUMP TO OPERATION 2 .
14 TEST PRODUCT-NO (B) AGAINST ; IF EQUAL GO TO OPERATION 16 ;
   OTHERWISE GO TO OPERATION 15 .
15 REWIND B .
16 CLOSE-OUT FILES C ; D .
17 STOP . (END)
\end{verbatim}

It is common for students to struggle to discern the meaning of this program (updating prices of items in an product inventory) despite their substantial programming experience.
In experiencing this struggle, I highlight that the use of natural language syntax is no guarantee that programmers will find a language usable, and that usability can only be understood relative to a given societal context.
We use this opportunity to explain why the program makes extensive use of \verb|GO TO| and \verb|JUMP TO| operations --- structured programming had not yet been invented, yet is such a standard feature today that the use of \verb|GO TO| is by far the exception rather than the rule.
The familiarity of natural language does not override the unfamiliarity of pre-structured programming, nor did FLOW-MATIC's contribution of one usability feature (natural language syntax) override the many usability challenges that remained in languages of that era.

The reflection on this sample program then turns to teasing apart the different kinds of complexity that can arise in a PL syntax, distinguishing, e.g., a verbose syntax from a syntax which requires a large number of keywords or production rules.
We observe historical trends in how PL designs have engaged with both forms of complexity.
Over the years, verbose code has remained widespread, yet improvements in editor automation have substantially reduced the number of user actions required to edit verbose programs, automation not available in the FLOW-MATIC era.
In contrast, we lack clear examples of automation for overcoming complexity of formal grammars or high keyword counts, and have instead seen languages with lower keyword counts take over in mainstream use. This observation is intrinsically linked with guiding principle of natural language, as PLs which seek to remain fully faithful to a natural language syntax tend to have higher keyword counts and production rule counts than others.
Instead, we have been left largely with languages which pull their keyword vocabularies from natural language, but make no attempt at grammaticality.

%\subsection{Inform}
%\label{sec:cs-inform}

\subsection{Processing}
\label{sec:cs-processing}
Processing~\cite{reas2006processing} is a language intended for use by media artists, such as visual, audio, and video art forms, and which has also seen use in pedagogy applications.
The Processing lesson comes after Natural Language.
It reinforces several key themes from the prior lecture: that the work of defining an audience has concrete impacts on the designs of actually-existing PLs, that this design is culturally and historically contingent, and that textual analysis of the written historical record is a means for unpacking these contingencies.
The differences between Processing and FLOW-MATIC are equally essential: most notably, Processing remains in widespread introductory-level use as of this writing, and is far more likely to be received as ``familiar'' by contemporary readers.
It is this contrast between unfamiliar and familiar which helps students internalize the idea of language designs being located in specific times in history.
In contrast to FLOW-MATIC, free web-based implementations of Processing are widely available, allowing students to experiment with Processing code in class, bringing a moment of personal experience to their analysis of the language.

The course's use of \emph{multiple} case studies with recurring themes is itself important, because this creates space to engage with underlying ideas of methodology.
The foundational lectures in design provide a primer on interpreting survey responses through the widespread qualitative method  of thematic analysis~\cite{braun2012thematic};
the FLOW-MATIC and Processing case studies both reveal that the qualitative paradigm includes far more than just the analysis of survey data.
The main ``facts'' presented to students in the Processing lesson are a list of themes, specifically PL design values, reconstructed from an informal thematic analysis process.
Compared to the FLOW-MATIC case study, the Processing lesson in particular highlights the diversity of sources used to analyze the themes.
The analysis is based on academic texts produced by Processing designers, the lived experience of the lesson author and students in writing programs, and oral communication with other Processing programmers.
In contrast to FLOW-MATIC, an analysis of Processing has ready access to primary sources beyond the written record.

The themes provided for the analysis of Processing are \emph{visuality}, \emph{immediacy}, and \emph{continuity}, the latter of which is explained in relation to the basic historical concept of \emph{contingency}.

To observe \emph{visuality} as a theme of Processing may at first seem thoroughly non-deep, contradicting the thesis of this paper.
On the contrary, the depth lies in the discussions that fall out of the themes.
I give one small example of how the brief hands-on experiences of programming in a classroom can add nuance to these themes.
One student remarked that ce\footnote{This student's pronouns are ce/cer/cers. These are \emph{neopronouns}, i.e., neologisms used by choice as personal pronouns, often for purposes of self-expression.} found \verb|printf|-style debugging much harder in Processing than other languages ce knew, perhaps because of its focus on visuals.
This quick remark complicates the common assumption that visual languages are automatically easier to use than textual ones; this relation may flip if the user is primarily comfortable with text.
We align this discussion of visuality with the core idea of PLs as interfaces.
In drawing this connection, I further contrast Processing, where visuals are a core program \emph{output}, with languages whose syntax is visual or whose inputs are images.
The other aspect of depth lies in contrasting visuality from the other themes, such as immediacy.

We characterize \emph{immediacy} as the experiential goal that when a programmer makes a change to a program (and attempts to re-run it), there is no perceptible delay between making the change and observing its impact on the behavior of the program.
We align this value with the broader principle of self-efficacy and the idea that programmer motivation is tied to self-belief in the capacity to plan and execute  programming tasks.
Immediate feedback as provided in Processing, by shortening the iterative development cycle of a program, may (be intended to) provide an increased sense of control.
We contrast several concepts with which immediacy is sometimes associated but from which it is distinct. In particular, Processing's immediate feedback is visual in nature, but immediacy is by no means restricted to visually-oriented programming environments, with interactive programming prompts (REPLs) being a longstanding PL feature.
Likewise, we can separate the question of immediacy from the question of typing discipline.
Many of the best-known REPLs are for dynamically-typed languages such as Python and various Lisps; because dynamic typing reduces the number of potential error messages between the writing and first execution of a program, it can be seen as a form of immediacy support.
Yet this does not remove the fact that many languages with robust (e.g. Hindley-Milner-style) type systems such as Haskell and Standard ML also provide well-established REPL interfaces. The decisions of typing discipline and immediacy are neither equivalent nor fully independent.

\emph{Contingency} is the basic concept that the occurrence of each historical event is dependent on the occurrence of previous events, and thus not inevitable, i.e., history could have unfolded differently.
In the HCPL context, contingency is presented as the concept that if past PLs had been designed or employed differently, the design decisions of new PLs might have been made differently.
The concept of contigency serves as the basis of two potential opposing design values: \emph{continuity} vs.\ \emph{radicalism}. 
Continuity is defined as the value that pre-existing design decisions are preferred by default until a compelling argument for a change is presented, while radicalism is the opposite value that novel designs are preferred by default.
Examples of radicalism can be found in \emph{esoteric programming languages}; Processing in contrast emphasizes continuity.
Continuity is closely linked to the notion of \emph{familiarity}: by identifying the specific user audience of a specific design, one can assess which existing design decisions are most likely familiar to the given audience, and thus decide the baseline set of design decisions against which an analysis of continuity or radicalism will be made.

\begin{figure}
\centering
\includegraphics[width=4in]{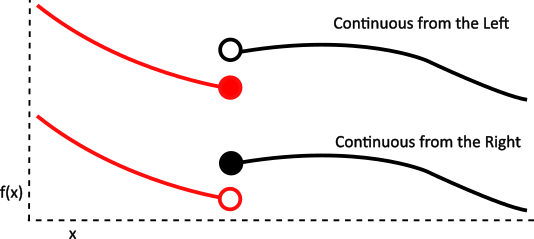}
\caption{Diagram of left-continuous and right-continuous mathematical functions}
\label{fig:continuity}
\end{figure}

The notion of continuity generalizes the notion of familiarity by employing a metaphor from a metaphor from elementary calculus: a function of a single variable could be \emph{continuous from the left}, \emph{continuous from the right}, both, or neither.
Such mathematical functions are depicted in \rref{fig:continuity} and used as a visual metaphor in the course.
Continuity from the left means that a PL is continuous with respect to the design decisions with which the audience is familiar before learning to use it.
Continuity from the right means that a PL is continuous with respect to the design decisions which the audience is likely to encounter \emph{after} learning to use it.
This theme fell out of analyzing the Processing designers' writings about their pedagogical uses of the language, synthesized with the present author's experiences using the languages Racket and Standard ML in the classroom.
Processing intentionally employs many syntax choices that align closely with widespread languages such as C and JavaScript, but the design behind this decision was not so simple as an appeal to languages that students already know.
In contrast, Processing is often used with \emph{first-time} programmers, who thus have no pre-existing attachment to a particular syntax. Instead, it is is designed with the recognition that some first-time students, even those who start from media backgrounds, will pursue further computer science studies, and then \emph{when they do so, they ought to find Processing relevant to future studies}.
The Processing strategy is not to make this language popular in industrial use, but to make it so visually similar to popular languages that students can immediately recognize programming concepts learned through Processing as relevant to their future programming work.

In my own experience using languages specially designed for pedagogy (the \emph{How to Design Programs}~\cite{felleisen2018design} languages), it was a common refrain among students that the languages are not useful in their lives or, when exaggerated, ``not real code''.
Through this lens of continuity, we can reflect as educators on the breadth of strategies available to us. 
The author has certainly advocated to students that programming concepts learned in a teaching language are transferable, or that there is intrinsic value in learning a broad variety of PLs, yet is worth having multiple strategies at hand, and designers in the pedagogy space can use the language of continuity to explore how students might integrate their knowledge from introductory courses with later study.
Indeed, in the discussion of Processing with students, I have also pointed out recent curriculum which use specialized languages for teaching yet put explicit thought into the question of transitioning into new languages afterward~\cite{dcic}.

%\subsection{Penrose}
%\label{sec:cs-penrose}

\subsection{Twine}
\label{sec:cs-twine}
As PL educators, we often seek to convince students that foundational concepts will have concrete benefits in applied development work.
In the functional programming context, foundational concepts could include higher-order functions, functional transformations on persistent data, metaprogramming, and type safety.
Though HCPL is not solely focused on functional programs, type safety is a shared core theme, from which we reach the Theorist's core theme of seeking universal, guaranteed predictions about the behavior of a program.
This theme is particularly challenging to drive home in the classroom.
Often, we drive the theme home by writing specific programs in specific languages with strong guarantees, yet this has a limitation: students may conflate the details of a chosen language with the underlying principles. 
For example, languages with strong guarantees have sometimes scared students away with unfamiliar syntax.
Yet it is hard to do better than this approach.
As much as we wish we could show students the joy of inventing a new type system guarantee for the first time, this is typically a research-level task and thus out of scope.

In HCPL, we supplement the use of an advanced type system with a second approach:
walking students through the discovery of theoretical foundations in a programming tool that was never intended to have them.
Though the students are walked through a specific example, the application domain is far-removed enough from the foundations that the joy of discovering a new application is still captured.

Twine~\cite{Twine} is a graphical tool for building interactive fiction games.
Although it allows inserting code in languages such as JavaScript, it does not advertise itself as a PL.
Our case study on Twine shows that by looking through the lens of the Theorist, we can quickly see a new formal language and formal guarantees where none existed before.

A Twine game consists of a graph $G = (V,E),$ where vertex set $V$ is a set of passages of text and edge set $E$ captures transitions between passages.
The Twine game engine shows a single passage $v = (n,p)$ of text on the screen at a time, where $n$ is the title of the passage and $p$ is its text.
A player plays a game by clicking \emph{links} embedded in the passage, each of which brings up a new passage on the screen.
Each link $e$ can be modeled as $e = (u, v, \ell)$ where $u$ is the name of the passage where the link appears, $v$ is the name of the destination, and $\ell$ is the highlighted text on which the link appears within the passage\footnote{Because the same text might appear multiple times in the passage, this is not how a production-grade implementation of Twine works, yet it is a useful simplification for presentation purposes.}.
We assume that for a given pair $u, \ell,$ the value $v$ is unique.
Let $E$ stand for the set of all such edges, then a Twine game $G$ comprises $G = (V,E)$ and the current state $s$ of a game at any moment is captured by a name $s = n$ such that $(n, p) \in V$ for some $p$.

To the Theorist, these definitions scream out ``I am a graph!''
The labels of this graph represent a single input action out of an input sequence, suspiciously similar to the workings of a deterministic finite automaton (DFA).
At this point, the lesson turns to explore a more precise question: ``How does Twine relate to a DFA and what guarantees will follow?''
The fundamental idea is that the winning plays (input sequences) for a Twine game will correspond exactly to the accepted strings (language) of a DFA.

Recall that a DFA is defined as a tuple $(Q, \Sigma, \delta, q_0, F)$ where $Q$ is the set of possible states, $\Sigma$ the set of input actions, $\delta$ a function describing all possible state transitions, $q_0$ the inital state and $F$ the set of desired final states (accepting states).
The HCPL curriculum includes review of the foundations of regular and context-free languages, preparing students to explore Twine as DFAs.
Only $q_0$ and $F$ are not inferable from the graph structure $G$; we treat these values as parameters which would be supplied in addition to $G$.
This is a small assumption because every Twine game does specify a specific start passage and many Twine game designers consider some end states to be ``good''  and others ``bad.''
We define $Q, \Sigma, \delta$ in terms of $G,$ recalling $\delta : Q \times \Sigma \rightarrow Q$:
\begin{align*}
Q &= \{n~|~\exists p.~(n,p) \in V \} \cup \{\textsf{err}\}\\
\Sigma &= \{\ell~|~ \exists u,v.~ (u,v,\ell) \in E \}\\
\delta(q,\ell) &= \text{the unique }q'\text{ s.t.\  }(q,q',\ell) \in E, \text{ else \textsf{err}} 
\end{align*}
where \textsf{err} is a distinguished error state not appearing in $G$.

Having completed this definition, we introduce a core methodological concept: ``Once you notice that language X matches foundation Y, review what facts are known about Y and attempt to apply them to X.''
We apply this approach using two of the most fundamental facts about DFAs: 1) regular languages are the languages accepted by DFAs and 2) the pumping lemma characterizes the expressive power of such languages.
Recall the pumping lemma:
\begin{lemma}
Let $m$ be a DFA $(Q, \Sigma, \delta, q_0, F)$ and let $n = |Q|$ be the size of $Q$.
Let $s \in \Sigma^*$ be any string whose characters are elements of $\Sigma$ and let the length $|s|$ of $s$ be at least $n$.

Then $s$ can be decomposed into three consecutive subsequences $s_1s_2s_3$ such that $|s_2| \geq 1, |s_1s_2| \leq n$ and for all natural numbers $k,$ we have that $s_1s_2^ks_3$ is accepted by the DFA.
\end{lemma}

The pumping lemma directly allows inferring upper and lower bounds on the length of accepted strings; in Twine, this immediately yields bounds on the lengths of winning plays:

\begin{theorem}
The number of moves in the shortest winning play, if any exists, is less than the number of passages in the game.
\end{theorem}

\begin{theorem}
Let $n$ be the number of passages.
If there exists a winning play with $n$ or more moves, there exist arbitrarily long winning plays.
\end{theorem}

On an even more fundamental level, the unpacking the definition of regular languages (as the languages of DFAs) gives an exact characterization of the expressive power of Twine games:
\begin{theorem}
For every Twine game $G,$ there exists a regular expression which accepts exactly the winning plays of $G$.
\end{theorem}

These guarantees are tangible enough to ``feel real'' to students, yet have a close  tie with a familiar foundation (DFAs) to feel ``theoretical'' at the same time, supporting the goal of convincing students to invest in theoretical thinking.

The full lesson also develops an operational semantics for Twine, in order to reinforce prior lessons on the use of inference rules, yet the core takeaway holds with or without operational semantics.

%\subsection{Process Calculus}
%\label{sec:cs-pc}

\subsection{Torino}
\label{sec:cs-torino}

This case study uses the PL Torino~\cite{DBLP:journals/hhci/MorrisonVTATSCS20} to illustrate how concepts from disability studies can inform PL design.
% spiel2020nothing
Designing for disability and accessibility has a rich scholarly tradition which we will not try to recount in detail.
We merely note that mainstream design research has not always centered critical perspectives from disability studies nor the lived experiences of disabled people; for a selection of research works which do, we refer readers to a recent workshop from the conference CHI~\cite{spiel2020nothing}.

This lesson emphasizes the principles of disability as a spectrum, the tension between visibility and invisibility of disabled status, and to a lesser extent, self-determination of disabled people.
Lesson objectives regarding the disability spectrum include assessing PL design decisions for their usability for programmers whose disability symptoms vary substantially from day to day, or which are used on teams of programmers with differing disability levels.
The lesson objectives for visibility include identifying PL design decisions which force a disabled person to be visible or invisible vs.\ those which give the person a choice.

The  case study on Torino~\cite{DBLP:journals/hhci/MorrisonVTATSCS20}, supports these learning objectives.
A key way it supports the objectives is by highlighting that the notion of disability as a spectrum had fundamental impacts both on the syntax of Torino and on how its designers structured their user studies.
It does not directly support the learning objectives about self-determination; those objectives are supported through brief case studies about the research works and life experiences of disabled researchers in the PL community.

Torino was a research project which developed a PL for use by visually-disabled children ages 7--11 in a classroom environment where the level of vision varies substantially between students.
As with other PLs targeting this age range, it is not the goal to teach students to write full-fledged applications.
Instead, typical goals for this group range from understanding of computational thinking principles to promoting interest and sense of belonging in computing, with an eye to robustifying the CS education pipeline.
A major challenge in teaching this age group is their limited literacy, which makes (e.g., block-based) visual PLs a standard approach for reaching them. 
This is the fundamental challenge Torino addresses: if visual programming is the standard for children's PLs, what is a designer to do when vision cannot be relied on?

Torino's solution to this challenge is a \emph{tactile}, i.e.\ touch-based, syntax.
The block-based paradigm is transformed into a bead-based paradigm.
Every basic PL construct corresponds to distinct type of physical ``bead,'' each with its own shape that can be told apart by touch alone.
In contrast to virtual wires in block-based languages, physical wires are used to connect the input and output ports of each bead.
The beads include operations for playing and modifying sounds, as opposed to creating visuals.
The above description illustrates the novel designs that emerge when we focus on accessible design, but does not demonstrate the lesson objectives.
We address these next.

Of the lesson objectives, Torino most directly speaks to our objectives regarding the disability spectrum.
It is a crucial design decision that in Torino, the different bead types have both a distinct shape and a distinct visual appearance, including different colors.
A majority of legally-blind people have some sight, thus accommodating the use of vision is not a tangential goal, but a central one.
On the contrary, designers must remember broader classroom goals for this age group: social and emotional development are key goals for all students in this range, and are often a heightened concern for children whose disability status may lead to exclusion in other spheres of life.
By creating a language which can use both touch and sight for interaction, the Torino designers support the goal of allowing fully blind and partially-sighted disabled children to play and work together, which is supportive of their social development.
At this point, the lesson turns to show that this understanding of the disability spectrum is not only fundamental to the design of the language syntax (a tactile syntax with visual elements), but also essential in designing user studies.
The user studies performed in the Torino project specifically explored how fully blind and partially-sighted children work together, e.g., collaborating on a single program. 
These studies showed that both groups used the language effectively, but in different ways, with fully blind children physically scanning the beads to identify points of interest and partially-sighted children relying on sight to locate specific beads.
This detail demonstrates that principles from disability studies help shape which questions a researcher or designer even asks, let alone the design solutions they arrive at.

The lesson then explores the tension between visibility and disability, i.e., the tension of whether to disclose one's disabled status to people in daily life.
Though Torino's designers do not discuss this issue explicitly, the discussion of the disability spectrum in Torino provides a strong foundation for this next objective.
In the same classroom, some students' disabilities may be more visible than others, and the classroom context is itself a key piece of the puzzle.
Compared to society at large, the classroom is likely a relative safe space where the adults in the room are educated and supportive regarding disability issues, and it may be a space where children are willing to let down their masks and be visible in their disability, an act which provides space for community-building and support, yet represents vulnerability.
The lesson segues into how the same underlying design principle, allowing multiple modes of interacting with a single program, actually supports the agency of disabled people in choosing whether to be visible, i.e., whether or not they choose to mimic the programming approaches used by abled programmers.
To support the final goal, the lesson closes with highlighting the work of a visually-disabled researcher on languages which support multiple (i.e., hybrid) syntaxes~\cite{andersen2020adding}.

\subsection{C-Plus-Equality}
\label{sec:cs-cpe}
\textbf{Content Notice:} This section discusses misogyny and transphobia.

The Gender lesson opens with a case study named C-Plus-Equality (C+=)~\cite{cpeRepo}.
In contrast to the other case studies of HCPL, this case study is brief and is purely motivational in nature. To my knowledge, it is also the only case study which has never had its  semantics formally defined nor had a production-quality implementation. This makes it no less important and no less deep. Before exploring the depth of this study, we must frame its history.

The language C+= is not a ``PL'' in the traditional sense.
Rather, it is a programming language design proposal, distributed in the form of a GitHub repository, containing a minimal prototype implementation using the C preprocessor, design documents, and example programs.
The repository lists its author as ``The Feminist Software Foundation,'' but has been attributed by online observers~\cite{cpeKnowYourMeme} to one or more misogynist trolls organized through the anonymous image board 4chan\footnote{4chan, like the other chan boards, is a forum for conversation on a wide variety of topics, but is arguably best-known for its role as an a organizing space for far-right social activists.}.
As of this writing, C+= is not new, with the repository dating to 2013.

Its visibility and influence have waxed and waned over time, however.
As of the writing of HCPL, simple web searches about the intersection of gender and PL design still return C+= as the top result, far before the 
% cite all of the following
relevant works of HCI scholars like Burnett~\cite{DBLP:journals/iwc/BurnettSMMBKPJ16},  artists like micha c\'{a}rdenas~\cite{disturbance} and Mez Breeze~\cite{mezangelle}, or even prominent scholarly discussions within the humanities, e.g., via the HASTAC~\cite{hastac} blog.
The latter example is especially noteworthy because online observers~\cite{cpeKnowYourMeme} have cited C+= as a \emph{reaction to} the HASTAC blog post~\cite{hastac}.
This anecdote about search results is used as the opening motivation to the Gender lesson, as a way of uniting students who might otherwise be divided in their beliefs about society. Not all of the author's students are feminists, but few would dispute that their feminist classmates deserve access to intellectually-sound curriculum, and thus even more-conservative students tend not to object to the lesson's presence.

The depth of this case study lies in the fact that our classroom analysis of the language relies on foundational perspectives techniques from the humanities, and cannot be properly understood using the perspective of any other archetype.
Specifically, we unpack this case study by starting with the prompt: ``Can a PL Spread Misogyny?'' and remarking on the nature of the question.
Few computer scientists have ever been asked a question of this nature, yet similar questions are routinely asked in \emph{media studies}, i.e., they are frequently asked about popular media such as songs, movies, TV shows, or novels.
A key insight  of this case study is that PL design documents, implementations, and code more broadly can all be understood as media: after all, they are used to express the creative message of an author to a given audience.

The (methodological) depth of this case study lies firmly in the humanities.
In analyzing PL design documents as media, CS methodologies are deeply insufficient.
We wish to analyze and interpret a small, related set of documents in detail: a close reading.
Drawing on this insight, the lesson turns to a standard technique from media studies: close reading.
The broad technique of close reading comes in many specific forms, among which our analysis situates itself as a \emph{hermeneutic inquiry} in the tradition of Schleiermacher~\cite{schleiermacher1998,smith1991hermeneutic}.

Hermeneutic inquiry arises originally from the interpretation of religious texts, and might be described as ``reading as a form of divination.''
This method grapples with a fundamental challenge in textual interpretation: though an author expresses their own intentions in a text, the process of interpreting those meanings is fundamentally a creative and thus subjective process.
Recognizing the creative aspect, the hermeneutic inquiry method does not aim for reproducible interpretations of a text.
Instead, it engages with the author's creativity through the interpreter's own creativity.
An emphasis is placed on an internal consistency of the interpretation, particularly a consistent interaction between the parts  and the whole of the text, with language playing a pivotal role~\cite{smith1991hermeneutic}.
In the HCPL classroom, these insights are presented at a high level, emphasizing that we will pick apart the details of a text with focus on the meanings of specific details.
We emphasize that we will employ subjectivity as we analyze the meanings and motivations behind these details.

We now undertake a close reading of the C+= design documents and prototype implementation in the tradition of hermeneutic inquiry, as is done in the course.
We reveal how the documents promote both misogynistic and transphobic messages and how those messages are ultimately inseparable from one another.
We quote several pieces of the text for analysis.

The prototype implementation consists of a series of C preprocessor macros performing simplistic keyword replacements; we list three example keywords which were used for classroom discussion.
\begin{verbatim}
#define privileged private
#define yell printf
#define social_construct class
\end{verbatim}
Though this snippet is devoid of interesting technical content, its content as media is substantial.
By placing these three definitions together, the snippet suggest these concepts are closely linked in the author's mind both with each other and the concept of a feminist.
This text suggests the author views feminists as likely to describe social categories as being constructed, e.g., drawing on Foucault or Butler, and that they are likely to discuss the topic of social privilege. 
More mundanely, the right-hand sides suggest an understanding of social categories as classes, private fields as privileged, and yelling as a form of I/O.
The crucial line, however, is the definition of \verb|yell|.

In placing these lines together, the author suggests that they understand discussions of privilege and social constructionism as a form of \emph{yelling}.
In the context of well-established tropes of online anti-feminist discourse, the (literal and figurative) keyword \verb|yell| unmistakably evokes the trope of the hysterical feminist, who blames every smallest life inconvenience on patriarchy. 
This trope serves to erase all real, justified claims of gender inequality by flattening all speech by feminists into the category of hysteria.

Three (non-consecutive) snippets of the primary design document are also used for in-class analysis, quoted here [emphasis in original]:

\begin{quote}
\textbf{Constants are not allowed}, as the idea of a lack of identity fluidity is problematic. $\ldots$ any numeric value is a variable, and is required to take on at least 2 values over the course of the program, or the inHERpreter will throw a \textbf{Trigger Warning}.

A number can be an integer or a double or a long if xir so identifies xirself.

Booleans are banned for imposing a binary view of true and false. 
C+= operates paralogically and transcends the trappings of Patriarchal binary logic.
\textbf{No means no, and yes could mean no as well.}
\end{quote}

The transphobia of the first snippet requires some analysis, but is hard to argue against.
The key term is \emph{identity fluidity}, which is a frequently-used means of describing gender identity for some LGBT+ people, particularly genderfluid people.
Juxtaposed against this, the following sentence is a clear metaphor for gender transition, with the requirement for taking on multiple values standing in for a requirement that a person take on multiple genders throughout life, i.e., a nonsensical requirement for all people to be transgender.
The reference to trigger warnings further solidifies the trope of the hysterical feminist.
By putting anti-trans and anti-feminist messages in the same paragraph, the author reveals their view of trans advocacy and feminist advocacy as fundamentally linked.

In the second snippet, the key words are \emph{xir} and \emph{xirself}.
The words \emph{xir} and \emph{xirself} are \emph{neopronouns}, i.e., neologisms chosen as personal pronouns, often for self-expression.
Pronouns are ascribed to people, not programs, and by ascribing a neopronoun to a program, the text implies that its author believes that the desire of some queer people to play with language for self-expression is as ludicrous as calling a program a person.
In my classroom, this concern is not an abstract one.
Recall that \rref{sec:cs-processing} recounts a classroom interaction with a student who has neopronouns (specifically, ce/cer/cers).
If my student encounted C+= in the wild, ce would rightfully interpret it as a personal attack.
By making space for criticism of C+= discourse in the classroom, I concretely demonstrate to cer that ce will receive protection from cer academic community, the PL community, when attacked using our own jargon.

In the third snippet, against this backdrop, it is hard to see the discussion of Booleans as anything other than a reference to the gender binary.
In its third sentence, \emph{no means no} is a widely-understood slogan for advocating against sexual violence.
Once again, the text links feminism and trans advocacy by juxtaposing the gender binary with references to sexual violence.

In summary, this proposal simultaneously contains many references to PL features that are laughable on their face to any experienced programmer, and consistently juxtaposes them will well-established concepts from feminism and especially transfeminism. 
In doing so, it seeks to build the association in its reader's mind that 
gender equality, and particularly trans equality, are equally laughable ideas.

This concludes the close reading performed in the lesson.
By taking time to close-read this text as a class, we not only motivate the rest of the lesson, but demonstrate to students that humanistic techniques can give them, as computer scientists, a new and practical dimension of understanding that complements the methodologies with which they are more familiar.
%also: parml, cfg, re, peg, rust

%\section{Classroom Experience Reports}
%\section{Lessons Learned}
\section{Conclusion}
\label{sec:conclusion}
This paper is a part of the \emph{Human-Centered Programming Languages} (HCPL) project to develop an open-access course for the interdisciplinary study of programming languages.
The project also includes an open-access textbook, course materials, and companion paper.
The \emph{humans} in HCPL represent different \emph{scholars} of PLs, through five archetypes with different intellectual traditions: the Practitioner, the Implementer, the Theorist, the Social Scientist, and the Humanist.
I call the HCPL course structure a \emph{tour-of-humans} as opposed to a \emph{tour-of-features} or \emph{tour-of-paradigms}, because it is organized around these archetypes with an overarching goal of leaving students able to read and converse with each academic community.
Though some students enter the classroom wondering how each academic tradition could fit into the topic of PL, just as many are glad for the opportunity to explore such connections for their first time.
This paper was dedicated to exploring the role of case studies in HCPL.
We gave detailed descriptions of a subset of the HCPL case studies: FLOW-MATIC, Processing, Twine, Torino, and C-Plus-Equality.

A key intellectual goal of this paper is to clearly separate the role of case studies in a \emph{tour-of-paradigms} from their role in a \emph{tour-of-humans}.
The former usage has been widely criticized for many years as lacking intellectual depth.
We thus arrive at the colloquial ``thesis'' of this paper, that PL case studies can be \emph{deep}.

I made this thesis more precise: I illustrated how case studies serve as essential tools for teaching foundational concepts from the social sciences and humanities and also illustrated how these foundational concepts have clear implications for PL design.
Educators of PL have long held interest in PL design, yet intellectually serious scholarship on design always needs techniques well-beyond the realm of CS, typically from both the social sciences and the humanities.
When we view students as students of design, case studies are deep because they reveal insights that could not be found with \emph{any other method}.
As a student of design, engagement with actually-existing historical designs is central, not optional.

The HCPL curriculum is based in a fundamental assumption, which not all scholars might agree with:
the foundations of PLs include the foundations of design in the social sciences and the humanities. 
This assumption has profound implications for pedagogy, of which the heavy use of case studies is but the surface.
The assumption that design foundations \emph{are} PL foundations radically alters our perspective of \emph{who our courses are for}.
I acknowledge this shift is a radical one, and that even sympathetic educators might sit with this an idea a long time before entertaining it in their own pedagogy.

Yet, I close by inviting the audience to sit with this idea for the precise reason that when we inspect who our courses are designed for, we are presented with a substantial opportunity for inclusive pedagogy.
Many PL and FP educators believe in our hearts that our subject is fundamental, that it can be brought to application across an incredible array of topics, and that it is a topic for everyone.
As curriculum designers, our work goes beyond making all people feel welcome in our classrooms.
By engaging with a wide range of \emph{scholarly traditions}, we may continue to broaden the range of students who find a deep and lasting connection with the topics we hold dear.

%\section{Bibliography}
\nocite{*}
\bibliographystyle{eptcs}
\bibliography{paper}
\end{document}